\documentclass[12pt,a4paper]{article} 
\usepackage[paper=a4paper,left=32.0mm,right=32.0mm,top=32.0mm,bottom=32.0mm]{geometry}

\usepackage{amsmath,amssymb,latexsym,amsfonts,url,dcolumn}
\usepackage[usenames]{color}
\usepackage[latin1]{inputenc}
\usepackage{natbib}                
\bibpunct{(}{)}{;}{a}{}{,}         
\usepackage{graphicx}
\usepackage{booktabs}
\usepackage{lscape}
\usepackage{eurosym}
\usepackage{enumerate}
\usepackage{graphicx,multirow,tabulary,tabularx,threeparttable,ragged2e,xmpmulti,tikz}
\usepackage[multiple]{footmisc}
\usepackage{comment}
\usepackage{fancyhdr}

\usepackage{array,multirow,threeparttable}
\usepackage{rotfloat}
\usepackage{caption, subcaption}
\usepackage[onehalfspacing]{setspace}
\begin{document}
	\title{The Behavioral Economics of Intrapersonal Conflict: A Critical Assessment}
	\author{Sebastian Kr\"ugel\thanks{Technical University of Munich, Richard-Wagner-Strasse 1, D-80333 Munich, Germany, phone: +49 89 907793 280, e-mail: sebastian.kruegel@tum.de} \and Matthias Uhl\thanks{Technical University of Munich, Richard-Wagner-Strasse 1, D-80333 Munich, Germany, phone: +49 89 907793 280, e-mail: m.uhl@tum.de}} 
	\date{\today}
	\maketitle

\thispagestyle{empty}

\onehalfspacing

\begin{abstract}
	Preferences often change---even in short time intervals---due to either the mere passage of time (present-biased preferences) or changes in environmental conditions (state-dependent preferences). On the basis of the empirical findings in the context of state-dependent preferences, we critically discuss the Aristotelian view of unitary decision makers in economics and urge a more Heraclitean perspective on human decision-making. We illustrate that the  conceptualization of preferences as present-biased or state-dependent has very different normative implications under the Aristotelian view, although both concepts are empirically hard to distinguish. This is highly problematic, as it renders almost any paternalistic intervention justifiable.
	\end{abstract}



\newpage

\section{Introduction}

According to standard economics, rational actors rank present choices according to preferences over the causal consequences of actions. Causality implies the lapse of time, and preferences may change between the times of choice and of consequence. Many years of research in economics and psychology have taught us that changes in preferences between these two points in time may be the rule rather than the exception. For instance, numerous studies on intertemporal decision-making demonstrated that people's choices might reverse because of the mere passage of time. Goods that people plan to choose when consumption is in the distant future are often systematically different from the goods that people actually choose as the future draws nearer. People's preferences often are time-inconsistent. They are ``present-biased'' \citep[][]{RabDo1999} in the sense that immediate consumption appears to be excessively overvalued.

However, \textit{present-biased preferences}\footnote{We use the term of \cite{RabDo1999} for the general phenomenon that people appear to overvalue immediate satisfaction. Another term for the same phenomenon is, for instance, (quasi-)hyperbolic discounting.} are not the only instance of changing tastes. Changes in environmental conditions might also affect people's preferences. This second source of changing tastes has received increasing attention in the more recent economic and psychological literature because many of those changes follow a systematic and thus predictable pattern \citep[see, e.g.,][for a general discussion of systematic fluctuations in tastes]{Loew1996, Loew2000}. For instance, people tend to prefer healthy (e.g., apples) over unhealthy (e.g., chocolate bars) food in a satiated state, but their preferences reverse in a hungry state \citep[][]{ReLe1998}. The value ascribed to durable cold-weather items (e.g., winter jackets) is higher on a cold than on a warm day \citep[][]{Conlin2007}, and the value attributed to certain types of cars (e.g., convertibles) is different on a sunny compared to a cloudy day \citep[][]{Busse2014}. Such preferences are called \textit{state-dependent preferences}, with different states being associated with different preferences. Whereas preferences under this concept vary with the underlying state (of nature), the concept of present-biased preferences captures the idea that preferences change with a mere movement along the time line. 

With changing preferences, rational actors are confronted with the problem of how to incorporate those changes into their rankings of alternatives when making their choices. Which preferences should a rational actor obey? Those at the time of the decision or those at the time of consumption when the consequences of the decision occur?

Whereas economists are usually concerned about the ``pretence of knowledge'' \citep[][]{Hayek1989} when it comes to social choices, this concern seems to be much less pronounced when it comes to individual decision-making. In stark contrast to the revealed-preference approach \citep{Samuelson1938,Samuelson1948}, many behavioral economists appear to be able to tell what people's relevant (i.e., true) preferences are even if choices differ between two points in time. Unfortunately, however, the economists' answers to this question can be very different, depending on the underlying concept of preferences. In the context of present-biased preferences, the standard belief in economics is that people's true preferences are those at the time of the decision, because preferences at the time of consumption are distorted by temptation or a tendency for immediate gratification \citep[see, e.g.,][]{St1955,PhPo1968,Poll1968,La1997,RabDo1999}. In the context of state-dependent preferences, on the contrary, the traditional view among economists is that the relevant preferences prevail at the time of consumption, not at the time of the decision \citep[see, e.g.,][]{ReLe1998,LoDoRa2003}.

Rational actors, who behave in accordance with the normative expectations of most economists, would thus have to take into account in their decisions whether a change in preferences between choice and consequence is caused by a movement along the time line or by a change of the state of nature. Truly rational actors may be able to do this, but to an outside observer, this is very difficult to tell.

The problem gets thorny when it comes to questions of consumer sovereignty and paternalistic interventions. In light of present-biased preferences, the benevolent economist is inclined to promote people's preferences at the time of the decision. In the case of state-dependent preferences, in contrast, the economist would side with people's preferences at the time of consumption. However, because it is difficult for an external advisor to judge what causes changes in preferences, partisanship for one set of preferences over another is problematic and perhaps primarily ideologically driven.

Although it may seem surprising at first that partisanship for a certain set of preferences in economics would be so arbitrary, there is a unifying element between both approaches to changes in preferences: the Aristotelian view on human decision-making that prevails in economics \citep{MolStev2001}. People are regarded as unitary decision makers, possibly coupled with various behavioral biases. Examples of self-commitment are interpreted exclusively as strategic interventions of rational actors against their own temptations in the context of present-biased preferences. The desire at the time of consumption is more of an impulse than a real preference. People with present-biased preferences are therefore ``naive'' \citep[][]{RabDo1999} if they fail to suppress their future temptations due to a lack of farsightedness, and they are ``sophisticated'' \citep[][]{RabDo1999} if they use self-commitment measures to preserve their current preferences against upcoming changes of mind. Ulysses is the ideal conception of rationality here. For people with state-dependent preferences, the opposite is true. Self-commitment measures are out of the question. People with state-dependent preferences are sophisticated if they abandon their current preferences in favor of their future ones and somewhat naive if they fail to please their future preferences.

In the present article, we take a stance against the Aristotelian view on human decision-making that is inherent to the economic literature on present-biased and state-dependent preferences alike. Clearly, if the usual assumption of unitary decision makers is complemented with the attribution of behavioral biases, the Aristotelian view makes paternalistic interventions into the self-management of people appear attractive, because each person consists of only one true rationale (or telos) that is to be promoted. However, since the idea of a person's true rationale differs between present-biased and state-dependent preferences, almost any paternalistic intervention is justifiable. In many real-life cases, the underlying reason for a preference change is hardly detectable by external advisors. 

An alternative perspective that does not deify the assumption of a unitary decision maker and instead takes the notion of multiple selves and intrapersonal conflict seriously would likely lead to a more liberal and less invasive stance towards a person's self-management. So far, the idea of multiple selves has been discussed mainly in the context of present-biased preferences \citep[see, e.g.,][]{St1955, Thaler1981, Parfit1984, RabDo1999, jamison2010}. In this article, we will focus on an understanding of multiple selves in relation to the notion of state-dependent preferences. Prima facie, the state-dependency view of preferences could be more open-minded towards an interpretation of decision makers in terms of multiple selves, because it does not degrade a particular preference to an impulse. However, based on the Aristotelian view, it upholds the assumption that the preference at the time of consumption is the only preference that counts. According to this view, people with state-dependent preferences whose choices are not fully in line with their \textit{own} future preferences must be mistaken in predicting them  \citep[see, e.g.,][]{ReLe1998, LoDoRa2003, Conlin2007, SaLoGrBl2008, Simon2009, Busse2014, BuchKol2016}. It is not an intrapersonal conflict between multiple selves that is observed, but a unitary decision maker's inability to properly foresee what she wants.

In the following section, we illustrate some exemplary experiments and field studies on state-dependent preferences and outline what has been inferred from the results. In section 3, we offer an alternative interpretation of the results and discuss what would be necessary to distinguish empirically between the two interpretations. In section 4, we critically assess the Aristotelian view on individual decision-making with regard to state-dependent preferences at a conceptual level and illustrate that some implausible assumptions are necessary to defend the approach. In the final section, we conclude and outline the different philosophical and political implications of the two approaches presented for our understanding of self-management.

\section{State-Dependent Preferences: Empirical Findings and Interpretations}

For many real-life decisions, the temporal separation between choice and consequence amounts to several days, weeks or even months and longer. This is the case when we preorder lunch in the cafeteria for the next day, buy groceries for the upcoming weekend, order goods via the Internet or a catalog or book a vacation trip for the upcoming summer. Choices with a marked time lag between the decision and consumption are sometimes termed \textit{advance choices} and have been frequently used to study state-dependent preferences. The greater time lag as such, however, is not of particular interest, but only a means to an end. The crucial element in those studies is that the preferences---i.e., the \textit{states}---differ between the two points in time.

In laboratory or field experiments, state-dependent preferences were investigated in various situations involving, for instance, hunger \citep{ReLe1998}, cigarette craving \citep{SaLoGrBl2008}, sexual arousal \citep{ArLo2006} or pain \citep{ReadLoew1999}. Most of these studies employed a very similar procedure. The participants made an advance choice in period 1 for a predefined future period 2 with corresponding state $s_2$. In period 1, some of the participants were in the same state that they occupied in period 2 (i.e., $s_1=s_2$), while other participants were in a different state (i.e., $s'_1 \neq s_2$). The participants were led to believe that their choice in period 1 would count but when period 2 arrived, they could in fact remake their choice. The typical results of those studies were that the participants made systematically different advance choices in state $s'_1$ compared to in state $s_1$ and revised their choices more often in period 2 if the advance choice was made in state $s'_1$. 

Consider, for instance, the classical study of \citet{ReLe1998}. They conducted a field experiment in which office workers made an advance choice between healthy (e.g., apples) and unhealthy (e.g., chocolate bars) snacks, which they received at a designated time one week later when they were either hungry or satiated. Half of the participants made their advance choice in a hungry state, while the other half chose in a satiated state. Directly before the participants received their chosen snack at the designated time one week later, they were asked to remake their choice, but they did not know about this option at the time of their advance choice. \citet{ReLe1998} found that advance choices were indeed influenced by anticipated future hunger levels. People who expected to be hungry the next week chose unhealthy snacks more often than people who expected to be satiated. However, advance choices were also affected by hunger levels at the time of the choice. People who were hungry at the time of the choice chose unhealthy snacks more often than those who were satiated. \citet{ReLe1998} therefore concluded that people erroneously projected their current preferences onto the future.

Field data studies on state-dependent preferences are not as common, but they exist as well. Most of these studies have utilized certain weather data to show that weather conditions systematically influence people's advance choices in a way that is inconsistent with expected utility theory. For instance, \cite{Conlin2007} investigated catalog orders of cold-weather items and found that the colder the weather on the order date, the more likely a return of that item was once it was received. \cite{BuchKol2016} analyzed advance ticket sales for an outdoor movie theater and found that good weather (i.e., based on sunshine duration) increased advance ticket sales even though the weather at the time of the purchase did not predict the weather on the day of the movie. \cite{Busse2014} studied a large data set of vehicle transactions and found that buying decisions for convertibles or four-wheel drives were affected considerably by the weather conditions on the day of the purchase. Lastly, \cite{Simon2009} even found that certain weather conditions (i.e., cloud cover) on the day of the visit to an academically demanding university affected the enrollment decisions of prospective students regarding this university.

Clearly, all of these studies demonstrated that people display state-dependent \textit{behavior}. That is, people's advance choices appear to be overly influenced by their current preferences (or states). Based on the assumption in economics that the relevant preferences prevail at the time of consumption, a widely accepted interpretation of the observed behavior is that people systematically mispredict their future preferences. People wish to please their future preferences in a different state, but unfortunately, they are biased towards their current wants. This misprediction is sometimes referred to as an \textit{empathy gap} \citep[e.g.,][]{Loew1996, Loew2000, Loew1999} or \textit{projection bias} \citep[e.g.,][]{LoDoRa2003,Conlin2007,Busse2014}. The latter term underlines the view that current preferences are illegitimately projected to another state for which a decision is to be made. People seem to understand the direction in which their preferences will change but systematically underestimate the magnitude of this change \citep[][]{LoDoRa2003}.

In the framework of state-dependent preferences, projection bias can be introduced in the following way \citep[see][who first formalized the idea that people with state-dependent preferences systematically mispredict their future preferences]{LoDoRa2003}. Suppose a person's instantaneous utility of consumption in period $t$ is given by $u(c_t,s_t)$, where $c_t$ is the person's consumption in period $t$ and $s_t$ is the person's state in period $t$ capturing his or her preferences. Further suppose that this person is currently in period 1 with corresponding state $s_1$ and is trying to predict her future instantaneous utility from consuming $c_2$ in period 2 with corresponding state $s_2$ (where $s_1 \neq s_2$). This prediction is denoted $\tilde{u}(c_2,s_2|s_1)$. If this person has no projection bias, she will predict her future utility correctly. That is, her predicted utility will equal her true utility: $\tilde{u}(c_2,s_2|s_1) = u(c_2,s_2)$. If, on the other hand, this person is exposed to a projection bias, as described in \cite{LoDoRa2003}, she will understand the qualitative direction of the taste change but underestimate its magnitude. That is, her predicted utility will be somewhere in between her true future utility and her utility given her current state: $\tilde{u}(c_2,s_2|s_1) = (1 - \alpha) u(c_2,s_2) + \alpha u(c_2, s_1)$, with $\alpha \in \left[ 0,1 \right]$. A person with projection bias is a person with $\alpha > 0$, with the bias increasing with higher values of $\alpha$. Since a person with projection bias misperceives his or her future utility, such an individual may exhibit dynamic inconsistency even in the absence of present-biased preferences: she may make a systematically different choice in period 1 (for period 2) than if she were asked again in period 2.

Importantly, the projection-bias interpretation of state-dependent behavior is based entirely on mistaken beliefs. A person with projection bias behaves exactly as a rational person does with the one exception that she mispredicts her future preferences \citep[][]{LoDoRa2003}. Rationality here means maximizing intertemporal utility. 

\section{State-Dependent Behavior: An Alternative Interpretation}

Our main caveat with the interpretations of the empirical results is that the participants' intentions at the time of making their advance choices remain unknown. All that is observed is state-dependent behavior. The inference that this behavior results from systematic mispredictions of future preferences is based solely on the assumption that the relevant preferences prevail at the time of consumption and, more importantly, that people share this view. 

An alternative but rarely considered interpretation of the same behavior is that people do not mispredict their future preferences in a different state but rather disagree with them and therefore try to impose a ``better'' judgment on a later alter ego. A person may generally disapprove of candy in a satiated state but have a less negative attitude towards sweets in a hungry state (independent of the time of consumption). Likewise, a person may spend parts of her limited budget on a certain durable good in one state, whereas she would want to use the same money for a different durable good in another state, simply because her overall valuations diverge in both states. For instance, suppose a person can allocate parts of her budget to either a handcrafted surfboard or a snowboard. Suppose it is currently summer, but since the surfboard is custom-made and making it will take a few months, it cannot be used before the next summer. The person may still order the surfboard, knowing that she would decide differently during winter.

If people are aware of this disagreement and insist on their valuations at the time of the decision, they will attempt to impose a ``better'' judgment at the expense of a later alter ego. This imposition can be understood as a \textit{sympathy gap} or \textit{conflict of selves}. Whereas the empathy gap is an ineffective attempt to please preferences in a different state, the sympathy gap is an attempt to fight them. No previous study could potentially rule out the latter explanation. To the best of our knowledge, no previous study did even attempt to do that.

The sole purpose of the concept of state-dependent preferences lies in the possibility that peoples' ranking of the causal consequences of their choices might change. Many studies have shown that this is indeed the case and that changes happen often in a very systematic and predictable manner. However, the expectation that a person with a certain ranking of his or her choices is willing to ignore this ranking for a different one at the time of consumption does not strike us as self-evident. At the very least, this is an assumption that should be empirically validated before declaring state-dependent behavior a manifestation of biased beliefs about oneself.

The problem is that the assumption about state-dependent preferences is notoriously difficult to address empirically. To be able to observe whether people with state-dependent preferences honor their preferences at the time of the decision or those at the time of the consumption requires keeping state-dependent and present-biased preferences apart. Otherwise, a person's deliberate choice in favor of her preferences at the time of the decision over those at the time of consumption could always be an instance of sophistication in the context of present-biased preferences.

Conceptually, the distinction between state-dependent and present-biased preferences is not a very difficult task, but empirically it is so. Present-biased preferences operate on a movement along the time line, and state-dependent preferences on a change in states. However, state changes always come with a movement along the time line. While this is a nomological necessity, the passage of time is conceptually irrelevant for state-dependent preferences. Therefore, to test the underlying assumption about state-dependent preferences, it must be ensured that a possible imposition of current preferences on a later self is indeed due to a disagreement between states and not to the mere passage of time.

The difficulty of studying empirically the underlying assumption of human decision-making in the case of state-dependent preferences also illustrates the difficulty of attributing real examples of dynamic inconsistency to one source or another of preference change. State-dependent and present-biased preferences may easily coexist. Thus, a person's ranking of choices may vary with the underlying state, and on top of this, the person may be exposed to present bias as consumption nears. For an impartial, benevolent advisor who observes cases of dynamic inconsistency, it is therefore hardly possible to propose the ``right'' paternalistic measures---even under the Aristotelian view of unitary decision makers.

\section{Mispredicting the Future and Forgetting the Past: Projection Bias and the Depletion of Rationality}

In this section, we take the observation of state-dependent behavior for granted and critically assess the idea of projection bias on a conceptual level. Notice again that projection bias is based on the assumption that people view themselves as unitary decision makers who are sympathetic towards changes in their own preferences. In cases of state-induced preference changes, the ``true'' preferences appear at the time of consumption and it is these that have to be satisfied. Accordingly, advance choices that are systematically influenced by current preferences are a manifestation of systematic mispredictions---the projection bias. Inherent to this concept is the necessary condition that people are unaware of their current misprediction. If they knew that they would choose differently at the time of consumption, the supposedly misguided advance choice would not be the result of a misprediction of preferences. Instead, we would have to relinquish the Aristotelian view of unitary decision makers in this context of state-dependent preferences, because the choosing self apparently imposes its current preferences on the future self.

We intend not to argue here that projection bias does not exist at all but mainly to question its scope. Projection bias is usually thought to explain a wide range of phenomena, from everyday decision-making to suicide \citep[see, e.g.,][]{LoDoRa2003}. The potential scope of projection bias is explicitly not limited to once- or twice-in-a-lifetime experiences for which mispredictions may occur simply because of lacking experience. However, the projection-bias explanation of state-dependent behavior is less convincing in daily or recurrent decision situations, precisely because unawareness of the current misprediction is a necessary condition of the concept. In recurrent situations, people do not even have to remember \textit{how} they felt in a certain state in the past to become aware of their bias. They only need to realize \textit{that} they previously did not like the consequence of the same choice they are about to make again.

For instance, suppose a person orders a warm winter jacket on a cold day because she overestimates its future value and then returns the jacket upon delivery because the temperature increased and the person is no longer overestimating the value of the jacket, as suggested in \cite{Conlin2007}. So the person orders the jacket because it keeps her warm on cold days and returns the jacket for exactly the same reason. Could this person unconsciously overvalue and order winter clothing on another cold day just to realize a few days later that this was a mistake once again? Notice that it is not necessary for the person to understand how weather affects her valuation of winter clothing. She only needs to remember that she was split over the value of warm clothing in the past to realize that something is wrong. As soon as this happens, the unconscious misprediction becomes a conscious disagreement with her future self regarding the value of warm clothing. In our view, this certainly is conceivable, as the essence of individual decision-making is the weighing of trade-offs. Sometimes the personal ranking of goods goes in one direction, sometimes in another, and every now and then these rankings may be in conflict with each other.

For a decision situation to be recurrent, it is not necessary that iterative states be identical. A decision situation is recurrent if states at two different points in time are more similar to each other than they are to another state regarding the relevant criterion. For instance, suppose a person books her summer vacation trip during the winter and due to projection bias chooses an overly warm destination \citep[the example is borrowed from][]{LoDoRa2003}---a choice that the person regrets during the summer. The situation becomes recurrent if this person wants to book a summer vacation trip during the following winter again. It seems highly implausible that the person will remain completely unaware of her last summer's misery if she plans to choose an overly warm destination a second time. The corresponding states during summers do not have to be identical. It is sufficient that it is generally warmer in summer and she might get tired of the heat again, whatever the actual temperature will be in the upcoming summer.

The lack of learning in recurrent situations, which is necessary for the concept of projection bias, should not be mixed up with the apparent lack of learning in the context of present-biased preferences. Overvaluing immediate consumption or rewards appears to be largely innate and difficult to break off, even if people are aware of their inclination. Learning in the context of present-biased preferences typically refers to issues of self-control. Projection bias, however, is based entirely on mistaken beliefs. In recurrent situations, a projection-biased person not only holds false beliefs about herself in a future state but must additionally remain unaware of her repetitive mistake. As another illustrative example, suppose a person who is currently satiated falsely believes that she will prefer a salad to a meat at the next day's lunch. Therefore, she preorders the salad and finds out on the next day that she would have liked the meat for lunch. If she again preorders the salad instead of the meat on another day, the concept of projection bias requires her to have forgotten her previous disappointment and to mispredict her desires during lunch once again. Importantly, in the concept of projection bias, she preorders the salad because she believes at the time that this is what she \textit{wants} for lunch on the next day, not what she thinks on the day of choosing she \textit{should} have for lunch the day after.

Overall, the projection-bias interpretation of state-dependent behavior requires a degree of forgetfulness about (dis)satisfaction with past choices in recurrent situations that appears implausible for an otherwise fully rational person. At the very least, the assumption regarding forgetfulness should be an explicit part of the concept. In recurrent situations, it is not sufficient that people systematically mispredict their preferences in consumption-relevant states; they must also be forgetful with respect to past choices and corresponding experiences. This certainly raises the question of what remains of rationality if people act ``as if'' they choose the utility-maximizing consumption bundle, but unfortunately for the wrong set of preferences and with complete absentmindedness towards their own experiences.

\section{Concluding Remarks}

About forty years ago, Thomas \cite{Schelling1984b} proposed that we as economists admit ``not only unidirectional changes [of preferences] over time, but changes back and forth at intervals of years, months, weeks, days, hours, or even minutes, changes that can entail bilateral as well as unilateral strategy'' (p. 6). Schelling called these preferences ``alternating preferences'' and suggested understanding and modeling phenomena such as these as manifestations of multiple selves. Each of these selves is characterized by its own preferences, which may be identical in many respects. In some others, however, preferences may differ fundamentally so that it seems impossible to compare utility differences between these selves or even add up their utility collectively \citep[][]{Schelling1984b}. Schelling concluded that ``this phenomenon of rational strategic interaction among alternating preferences is a significant part of most people's decisions and welfare and cannot be left out of our account of the consumer'' \citep[][p. 5]{Schelling1984b}. Economists, Schelling noted, could of course ignore these phenomena, but if they do not, they should not have major difficulties in understanding decision makers as a collection of multiple selves \citep[][]{Schelling1984b}. After all, he considered the ``art of self-management'' as an integral part of the conditio humana \citep{Schelling1978}.

Several decades and numerous empirical studies later, we know that preferences can indeed fluctuate systematically, even in short time intervals. The phenomenon of alternating preferences or, in today's terms, recurrent situations of state-dependent preferences has become difficult to ignore. What in the interim seems to be somewhat forgotten is the idea that these preferences may not simply add up and may instead be expressions of different selves with their own rationality. The underlying similarity between Schelling's (1984b) anecdote about himself as a young boy who saw a movie about Admiral Byrd's first Antarctic expedition and subsequently decided to toughen himself against the cold\footnote{The anecdote goes as follows: ``As a boy I saw a movie about Admiral Byrd's first Antarctic expedition and was impressed that as a boy he had gone outdoors in shirtsleeves to toughen himself against the cold. I decided to toughen myself by removing one blanket from my bed. That decision to go to bed one blanket short was made by a warm boy; another boy awoke cold in the night, too cold to go look for a blanket and swearing to return it tomorrow. But the next bedtime it was the warm boy again, dreaming of Antarctica, who got to make the decision, and he always did it again'' \citep[][p. 8]{Schelling1984b}.} and the observation in \cite{Conlin2007} that people tend to return cold-weather items in warmer periods is utterly bizarre. The reasons for the relinquishment of warm blankets or clothes may have been different in both cases, but the fundamental problem is the same. For Schelling (1984b), it was an illustrative example of alternating selves, while for \cite{Conlin2007}, it was an instance of suboptimal behavior based on biased predictions of future preferences.

According to Spiegler's (2019) critique of behavioral economics, the interpretation of state-dependent behavior as a misprediction of one's own preferences and the resulting model of projection bias falls into the category of ``functional-form style'' modelling, where a standard economic model (here: intertemporal utility maximization with state-dependent preferences) is taken and the behavioral phenomenon is captured by an additional parameter. \cite{Spiegler2019} points out that this type of modelling may distort our understanding of the actual underlying problem and urges putting a stronger weight on a more ``conceptual framework,'' especially in the study of behavioral economic phenomena. Regarding our context, we fully agree with \cite{Spiegler2019}, because the question of which of people's changing preferences they themselves consider to be relevant is of fundamental importance for theorizing as well as providing policy advice.

\citet{MolStev2001} argue that the Aristotelian view prevailing in the social sciences treats persons as unitary decision makers, while the less popular Heraclitean view accepts that inner conflict is at the core of human existence. This latter view focuses on the processes and phenomena by which people resolve their internal conflicts. The concept of a person may be no less fuzzy than the concept of, say, a firm \citep[see also][]{Parfit1984}. \citet{Cowen1991} has noted that particularly influential literature on self-management bifurcates persons into a long-run, rational self and a short-run, impulsive or ``irrational'' self. Because it analyzes the strategies that the long-run self uses to induce cooperation from the impulsive self, he refers to this literature as the ``command view of self-management.'' Cowen argues that good self-management instead unleashes forces that enable personality growth. Giving up the asymmetry between the two selves represents a right step in the direction of a new view of self-management that deemphasizes control. 

The labels ``irrationality'' and ``bias'' \citep[see, e.g.,][]{BernRang2007} are important weapons in the new paternalists' campaign against the self-determined citizen. Behavioral paternalists frequently use the notion of multiple selves \citep[e.g.,][]{Thaler1981,BernRang2004}. In fact, they are dedicated Aristotelians in modeling torn persons as unitary decision makers who happen to have a hyperbolic discounting function or other biases. Originally, the Aristotelian view used to emphasize normative individualism---i.e., the dictum that you can't argue about taste \citep{StigBeck1977}. This changes dramatically when the view is complemented with the attribution of behavioral biases. The exponential discounting function, for instance, then becomes a person's telos and thus a normative benchmark. However, the behavioral paternalists' conception of rationality may be overly narrow \citep{RizzoWhitman2019}. Intrapersonal conflict may not be reducible to myopia whereby an impulsive self does not internalize the effect of its ``selfish'' behavior on future alter egos. On the contrary, the impulsive self may be the only self that understands the true magnificence of a blissful moment of indulgence \citep{Schelling1984a}.  The Heraclitean perspective casts doubts on the naturalness with which the new paternalists take sides.

\clearpage
\setstretch{1.1}
\bibliographystyle{ecta}
\bibliography{empathygap}

\end{document}